\documentstyle{article}
\newcommand{\hochpunkt}[1]{\mbox{$^{\raisebox{.3ex}{\scriptsize #1}}_{\raisebox{.6ex}{\hspace{.17em}.}}$}}
\newcommand{\hoch}[1]{\mbox{$^{\raisebox{.3ex}{\scriptsize #1}}$}}
\begin{document}

\begin{center}
{\LARGE \bf Cyclic variations with twice the accretion disk precession period
in the old nova V603~Aquilae}
\vspace{1cm}

{\Large \bf Albert Bruch}
\vspace{0.3cm}

Laborat\'orio Nacional de Astrof\'{i}sica, Rua Estados Unidos, 154, \\
CEP 37504-364, Itajub\'a - MG, Brazil
\vspace{0.5cm}

{\Large \bf Lewis M. Cook}
\vspace{0.3cm}

American Association of Variable Star Observers, 1739 Helix Ct. \\
Concord, CA 94518, USA
\vspace{1cm}

(Published in: New Astronomy, Vol.\ 63, p.\ 1 -- 5 (2018))

\vspace{1cm}
\end{center}

\begin{abstract}
A dense series of long and high time resolution light curves
of the old nova V603~Aql, covering 22 nights (19 of which are consecutive),
are analyzed in order to identify and characterize variations on the 
time scale of hours
and days. The well known 3.5 hour modulation, observed many times
in the past and considered to be due to a long lasting, albeit not entirely
stable superhump, is recovered at a period of $0.1453$ days and an 
amplitude of $\sim 0\hochpunkt{m}062$. Most interesting, 
however, is the detection of highly significant brightness variations with
an amplitude of $\sim 0\hochpunkt{m}0.050$ and
a period of 5.85 days which is to a very high precision 
equal to {\em twice} the beat period between the orbital and the
superhump period. The latter is generally interpreted as the precession period
of an eccentric accretion disk. The origin of these long term variations
remains unknown.
\vspace{1ex}

{\parindent0em Keywords:
Stars: novae, cataclysmic variables --
Stars: individual: V603 Aql}
\end{abstract}

\section{Introduction}
\label{Introduction}

Among all known classical novae V603~Aql has the highest apparent magnitude
in quiescence. As such, it is very well studied and the amount of literature 
on the object is vast. V603~Aql erupted in 1918 and reached quiescence in 
1937 (Strope et al. 2010). From then on it exhibited a slight secular fading 
to a mean quiescent magnitude of $\sim$11\hochpunkt{m}8 in recent years 
(Johnson et al. 2014). 

The orbital period was first determined spectroscopically by Kraft (1964),
later refined by Drechsel et al.\ (1982) and more recently by
Arenas et al.\ (2000). Peters \& Thorstensen (2006) were able to combine
their own radial velocity measurements with results from previous studies
to derive long term ephemeries, establishing the orbital period to be
0.13820103 days (3\hoch{h} 19\hoch{m} 0\hochpunkt{s}57). 
Arenas et al.\ (2000) measured component masses of $M_1 = 1.2 \pm 0.2\,
M_\odot$ and $M_2 = 0.29 \pm 0.04\, M_\odot$ and an orbital inclination of
$i = 13^{\rm o} \pm 2^{\rm o}$. 

V603~Aql exhibits a puzzling multitude of photometric periods none of which
is identical to the orbital period. Haefner \& Metz (1985) observed 
polarimetric variations with a period of 2\hoch{h} 48\hoch{m} and suggested
an intermediate polar model. However, Cropper (1986) did not detect
polarization in V603~Aql. Drechsel et al.\ (1983) found variations of the
x-ray flux which they considered to be compatible with the orbital period.
But this interpretation is not unique. Haefner \& Metz (1985) showed that 
these variations are also compatible with their polarimetric period. It is
even compatible with a 61\hoch{m} photometric period observed by Udalski \&
Schwarzenberg-Czerny (1989). Schwarzenberg-Czerny et al.\ (1992) also
found pulsations in the UV continuum radiation with a period very close to
the latter. Bruch (1991) could not confirm the 61\hoch{m} period
but found instead indications for a period close to 1\hochpunkt{h}5. All
these periods remain unconfirmed, and the credentials for V603~Aql as an 
intermediate polar remain doubtful\footnote{see Koji Mukai's Intermediate
Polars Home page (https://asd.gsfc.nasa.gov/Koji.Mukai/iphoe/iphome.html)}.

The only consistent modulation present in the light curve of
V603~Aql has a slightly variable period close to 3\hochpunkt{h}5, about
6\% longer than the orbital period. It has
first been detected by Haefner (1981) and was then confirmed by many authors
(Udalski \& Schwarzenberg-Czerny 1989, Bruch 1991, Patterson \& Richman 1991,
Patterson et al. 1993, Hollander et al. 1993). Additionally, Patterson et 
al.\ (1997) observed a modulation with a period being about 3\% less than
orbital period. Today, the longer of these  variations is interpreted as a
positive (because its period is longer than $P_{\rm orb}$) 
superhump in V603~Aql, while the shorter one is considered to be
an occasionally visible negative (period shorter than $P_{\rm orb}$) superhump. 

Superhumps\footnote{Unless explicitely stated otherwise, the term 
``superhump'' is used subsequently to indicate a positive superhump.}
are routinely observed in superoutbursts of SU~UMa type dwarf
novae. They are explained by tidal stresses in an accretion disk which
has become elliptical as a result of a 3:1 resonance between the rotation
period of matter in the outer accretion disk and the orbital period of
the secondary star in a cataclysmic variable (CV).
The superhump period $P_{\rm SH}$ is slightly
longer than the orbital period $P_{\rm orb}$ because of apsidal precession
of the elliptic disk. The superhump, orbital and precession periods are
related to each other by $1/P_{\rm prec} = 1/P_{\rm orb} - 1/P_{\rm SH}$.
Instead, negative superhumps are explained as variations of the brightness
of a warped accretion disk with a retrograde precession of the nodal line.

Although most common in short period dwarf novae in superoutburst, superhumps 
are also seen in an increasing number of longer period CVs, independent of
outbursts; mostly in old novae or novalike
variables where they can persist over long periods of time. V603~Aql is one 
of them. Others include
TT~Ari (Belova et al.\ 2013 and references therein; Smak 2013),
KR~Aur (Kozhevnikov 2007),
AT~Cnc (Nogami et al. 1999, Kozhevnikov 2004),
TV~Col (Retter et al. 2003),
V751~Cyg (Patterson et al.\ 2001, Papadaki et al.\ 2009),
V795~Her (Patterson \& Skillman 1994, Papadaki et al. 2006), and
V378~Peg (Kozhevnikov 2012). This list is not meant to be exhaustive!

Here, we take advantage of a particular long and densely sampled
series of light curves of V603~Aql in order to study the superhump
modulations as well as variations on the time scale of days with the purpose to
identify possible brightness changes on the disk precession period such as
has been claimed to be present in some CVs (see Table~4 of Yang et al.
2017). In Sect.~\ref{Observations}, we present these observation. Variations
on the superhump and on longer time scales are analysed in 
Sect.~\ref{Superhumps and long term variations} and then 
discussed in Sect.~\ref{Discussion}. A short summary of the results
concludes this paper in Sect.~\ref{Summary}.

\section{Observations}
\label{Observations}

The present study is based on light curves observed by one of us (LC)
during 22 nights between 2003, June 21, and July 18. A gap of seven
nights separates 19 light curves taken in subsequent nights from a shorter
set of 3 light curves, also taken in subsequent nights. All observations 
were obtained using the 73cm prime focus reflector at Concord observatory
which is part of the Center of Backyard Astrophysics
network\footnote{cbastro.org}. 
Integrations lasted 15\hoch{s}, using a commercial CCD camera 
(Genesis 16). Including overheads this resulted in a time resolution of
35\hoch{s}. With very few exceptions the individual light curves extend 
over 5 -- 6 hours or more. To our knowledge this is the densest and most
homogeneous set of light curves of V603~Aql ever studied.

No filters were employed during the observations.
Bruch (2018) found that his unfiltered CCD photometry was approximately 
equivalent to observations in the $V$ band. He used detectors
which are more sensitive in the blue than the one used here (KAF 1602E).
Thus, the effective wavelength $\lambda_{\rm eff}$ of the present light curves 
is probably somewhere between $V$ and $R$. Noting that there is only a small 
magnitudes difference between $V$ and $R$ in V603~Aql (we measured an average 
$V-R = 0.08$ in unpublished multicolour light curves) no significant 
uncertainties is introduced assuming that the data refer to the $V$ band.

All exposures were processed (dark subtracted, flat fielded) using commercial
software. Magnitudes were calculated by referring to the $V$ band magnitude
of the comparison star UCAC4 453-082428.  
Time was transformed into barycentric Julian Date on the Barycentric
Dynamical Time (TDB) scale using the online tool provided by 
Eastman et al.\ (2010). For subsequent data analysis the MIRA software system 
Bruch (1993) was employed. A journal of observation is given in Table~1.

\begin{table}
{\bf Table 1:} Journal of observations
\vspace{1em}

\begin{tabular}{lccc}
\hline
Date   & Start & Duration & No. of   \\
(2003) & (UT)  & (min)    & integrations \\
\hline
Jun 21 & 7:39 & 151 & 248 \\
Jun 22 & 5:13 & 357 & 585 \\
Jun 23 & 5:49 & 319 & 532 \\
Jun 24 & 5:53 & 312 & 442 \\
Jun 25 & 5:32 & 324 & 508 \\
Jun 26 & 5:05 & 380 & 477 \\
Jun 27 & 4:47 & 397 & 643 \\
Jun 28 & 5:01 & 382 & 598 \\
Jun 29 & 4:46 & 396 & 541 \\
Jun 30 & 5:46 & 332 & 487 \\
Jul 01 & 4:37 & 392 & 640 \\
Jul 02 & 5:08 & 358 & 572 \\
Jul 03 & 5:00 & 362 & 593 \\
Jul 04 & 4:44 & 367 & 582 \\
Jul 05 & 8:18 & \phantom{3}85 & 139 \\
Jul 06 & 4:47 & 378 & 623 \\
Jul 07 & 4:51 & 370 & 602 \\
Jul 08 & 4:56 & 361 & 615 \\
Jul 09 & 5:01 & 352 & 586 \\
Jul 16 & 4:59 & 338 & 575 \\
Jul 17 & 4:35 & 355 & 612 \\
Jul 18 & 4:33 & 352 & 621 \\

\hline
\end{tabular}
\end{table}

All light curves are available at the AAVSO International 
Database\footnote{https://www.aavso.org}.

\section{Superhumps and long term variations}
\label{Superhumps and long term variations}

\begin{figure}
\parbox[]{0.1cm}{\epsfxsize=14cm\epsfbox{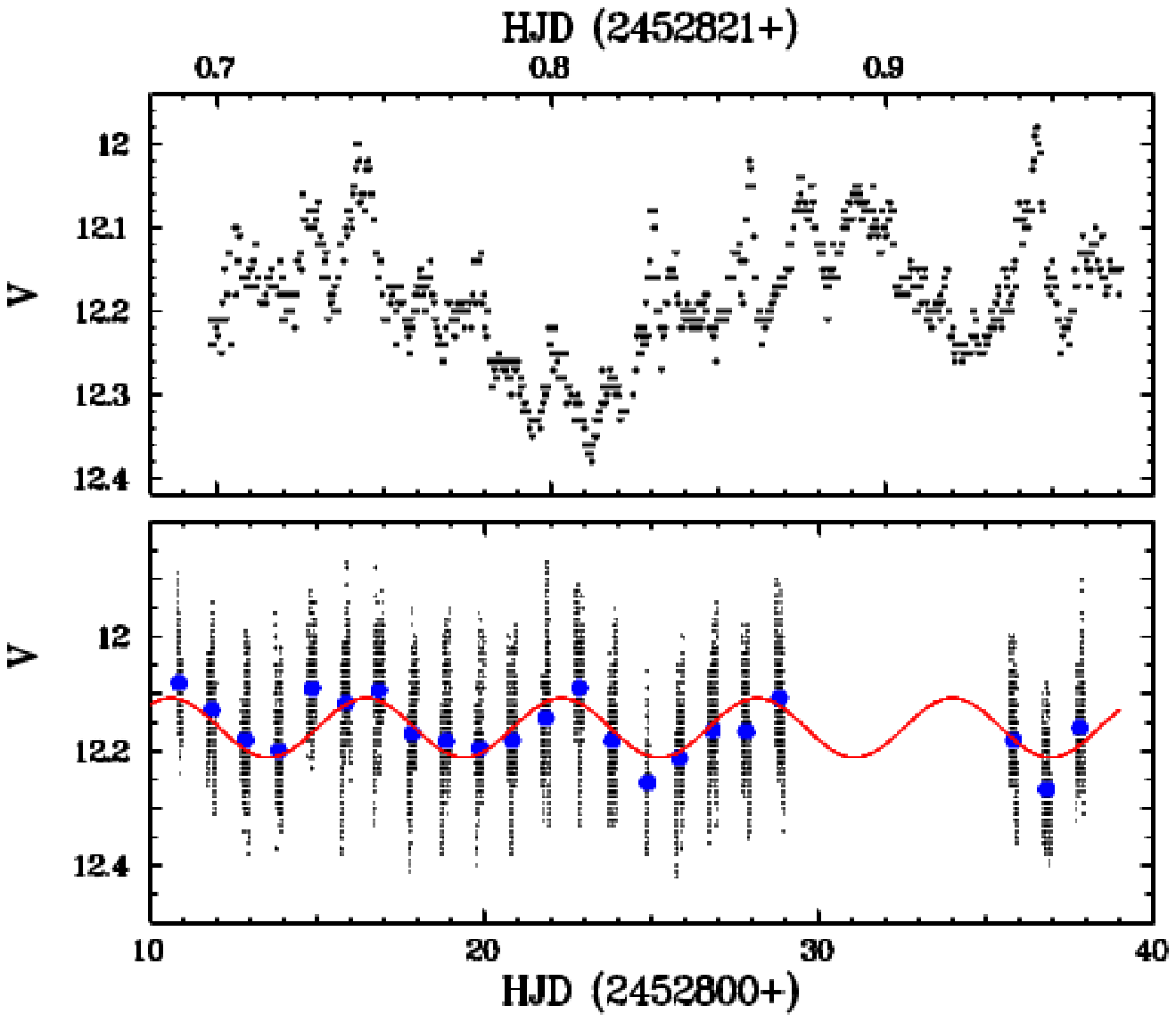}}
      \caption[]{{\it Top:} Light curve of V603~Aql of 2003, July 1.
                 {\it Bottom:} Combined light curves of V603~Aql observed
                 in 22 nights between 2003, Jun 21 and July 18. The average
                 nightly magnitudes are indicated by light blue dots. The red
                 curve represents a least squares sine fit with the period
                 fixed at 5.85 days.}
\label{lightcurves}
\end{figure}

As an example Fig.~\ref{lightcurves} (top) shows the light curve observed
on 2003, July 1. It is dominated by flickering activity typical for 
cataclysmic variables, but also exhibits a clear modulation on the time
scale of hours. In the lower frame of the figure the combined light curves
of the 22 nights during which V603~Aql was observed is shown. Variations
occurring during the individual nights are, of course, not resolved on this
scale, but a clear and apparently periodic modulation repeating on the time 
scale of several days is obvious, as indicated by the average nightly
magnitudes (light blue dots).

\input epsf

\begin{figure}
\parbox[]{0.1cm}{\epsfxsize=14cm\epsfbox{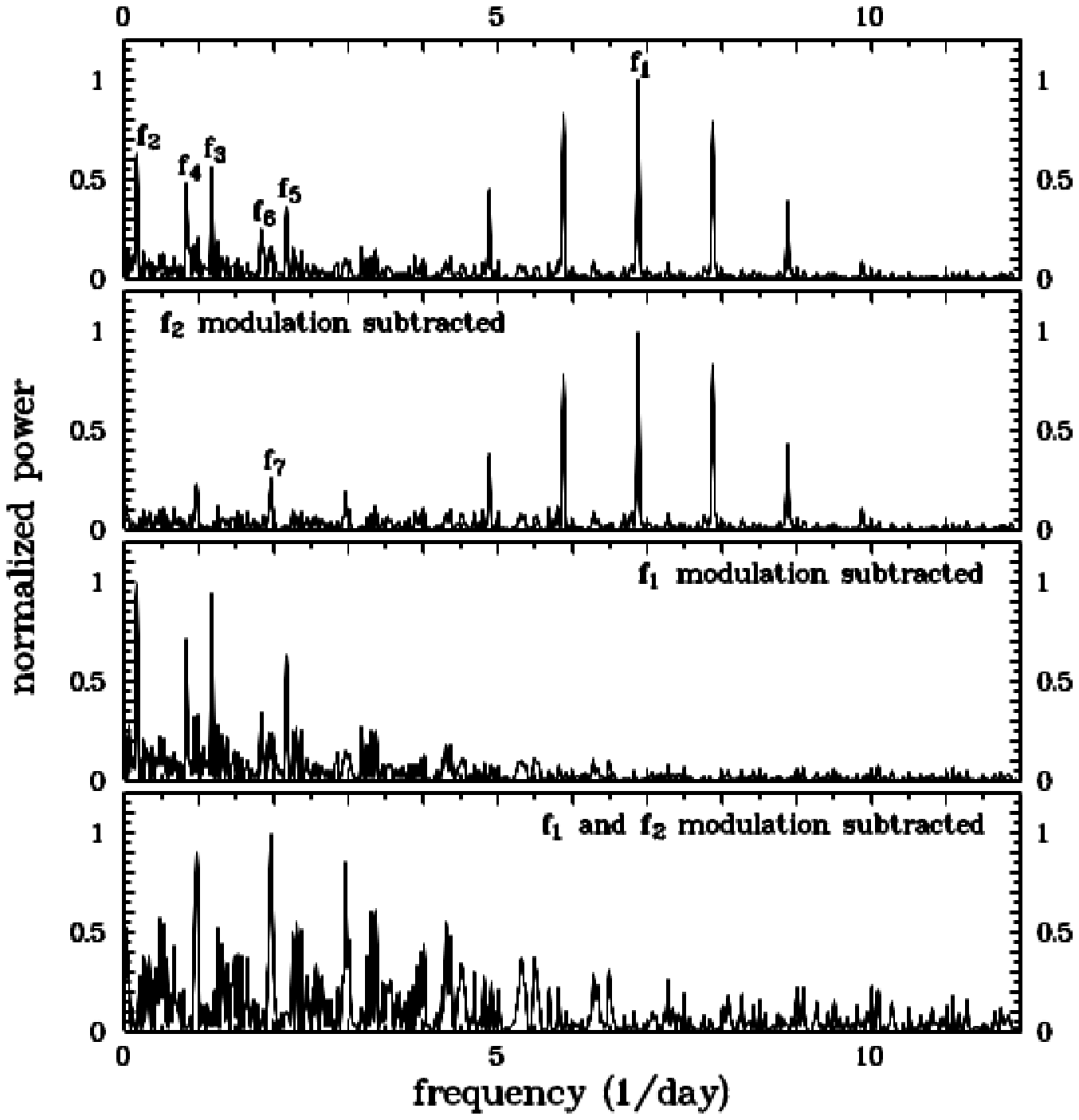}}
      \caption[]{{\it Upper frame:} Normalized power spectrum of the combined
                 light curves of V603~Aql observed in 22 nights between
                 2003, June 21 and July 18. Possibly significant peaks
                 are marked (disregaring the obvious 1/day alias peaks
                 around $f_1$). Their frequencies and periods are listed
                 in Table~2. {\it 2$^{nd}$ frame:} Power spectrum after 
                 pre-whitening of the data by subtracting a sine fit with 
                 the period fixed to $P_2$. {\it 3$^{rd}$ frame:} The same
                 after pre-whitening by subtracting a sine fit with the period
                 fixed to $P_1$. {\it Lower frame:} The same
                 after pre-whitening by subtracting a two-component sine fit 
                 with the periods fixed to $P_1$ and $P_2$.}
\label{powerspectra}
\end{figure}

Fig.~\ref{powerspectra} (top) shows the Lomb-Scargle periodogram
(Lomb 1976, Scargle 1982, hereafter referred to as power spectrum), 
normalized to the highest peak, of the combined light curves.
The plot is restricted to frequencies $<$12 cycles per day. No significant
signals were detected at higher frequencies. The power spectrum
is dominated by a maximum labelled $f_1$ in the figure. Disregarding the
symmetrical pattern of peaks at both sides of $f_1$, which are obvious 
1/day aliases caused by the data sampling, we identify several other peaks
labelled $f_2$ \ldots $f_6$ (in decreasing order of their height) in the 
power spectrum which appear to be
significant. The frequencies of the maxima of Gaussians fit to these peaks 
are summarized in Table~2 together with their corresponding periods $P_1$
\ldots $P_6$. Here, the error is conservatively defined as the width 
$\sigma$ of the best fit Gaussians.

\begin{table}
{\bf Table 2:} Significant frequencies and periods identified in the power 
spectrum of the combined light curves of V603~Aql
\vspace{1em}

\begin{tabular}{ccc}
\hline
No. & $f$ (1/day) & $P$ (days)  \\ 
\hline

1 & $6.881\phantom{0} \pm 0.014\phantom{0}$ & 
    $0.1453 \pm 0.0003$ \\ 
2 & $0.171\phantom{0} \pm 0.013\phantom{0}$ & 
    $5.85\phantom{00} \pm 0.43\phantom{00}$ \\ 
3 & $1.173\phantom{0} \pm 0.013\phantom{0}$ & 
    $0.853\phantom{0} \pm 0.009\phantom{0}$ \\ 
4 & $0.831\phantom{0} \pm 0.014\phantom{0}$ & 
    $1.203\phantom{0} \pm 0.021\phantom{0}$ \\ 
5 & $2.175\phantom{0} \pm 0.013\phantom{0}$ & 
    $0.460\phantom{0} \pm 0.003\phantom{0}$ \\ 
6 & $1.832\phantom{0} \pm 0.019\phantom{0}$ & 
    $0.546\phantom{0} \pm 0.006\phantom{0}$ \\ 
7 & $1.966\phantom{0} \pm 0.014\phantom{0}$ & 
    $0.509\phantom{0} \pm 0.004\phantom{0}$ \\
\hline
\end{tabular}
\end{table}

The period $P_1 = 0.1453 \pm 0.0003$~days can evidently be identified with 
the period of the superhump in V603~Aql (the 3\hochpunkt{m}5 modulation). 
It is very close to
that measured in previous observations by Haefner (1981) (0.144854 days),
Udalski \& Schwarzenberg-Czerny (1989) (0.14567 days), Bruch (1991)
(0.14468 days) and Patterson \& Richman (1991) (0.14548 days). The latter 
authors also comment on
seasonal variations of the period. The long term instability of the period
was confirmed by Patterson et al.\ (1993) who reported the period to drift
from 0.14602 days to 0.14663 days over the cause of about four months in 1991.

In order to verify, if the period over the time base of the current observation
is stable, power spectra of the first and second half of the entire
data set (eliminating the last three nights, separated by a 7 night gap from
the other light curves) were calculated. The resulting periods differ by 
19\hoch{s}, well within the formal error. Next, sections of three subsequent 
nights (with a two night overlap between adjacent sections) were investigated. 
No systematic drift of the period was
found. Thus, we conclude that over the time base of the present light curve
the 3\hochpunkt{h}5 period of V603~Aql remained stable within the achievable
accuracy.

The power spectrum also exhibits some peaks at low frequencies. To our 
knowledge, this frequency range has not been investigated in detail before. 
The strongest peak in this frequency range, $f_2$,
corresponds to a period of $P_2 = 5.85 \pm 0.43$ days. This modulation
is readily seen in the combined light curve. The red curve in the lower frame
of Fig.~\ref{lightcurves} represents a least squares sine fit with the period
fixed to $P_2$. The nature of these variations will be discussed below. 

In order to further investigate the properties of the various observed
frequencies in the combined light curve, we first pre-whitened the data
set by subtracting the $P_2$ modulation (i.e., subtracting the sine curve
in the lower frame of Fig.~\ref{lightcurves}). The power spectrum of the
result is shown in the second frame of Fig.~\ref{powerspectra}. The
frequencies $f_2$ -- $f_6$ completely vanished, indicating that $f_3$ --
$f_6$ are not independent from $f_2$. In fact, they can all be explained
as being caused by the window function which due to the data sampling has
a strong peak are $f_w = 1$. Within the error limits,
$f_3 = f_2 + f_w$, $f_5 = f_2 + 2f_w$, $f_4 = f_w - f_2$, and $f_6 = 2f_w - f_2$. 

However, apart from $f_1$ and its alias pattern the power spectrum now reveals
another system of fainter peaks. The strongest of these has a frequency of
$f_7 = 1.966 \pm 0.014$ per day (included in Table~2). The others are 1/day 
aliases. These peaks are also present in the original data set (before
pre-whitening) but dwarfed by stronger signals at neighbouring frequencies.
A first suspicion they they are somehow caused by the seven day gap close to
the end of the combined light curve cannot be confirmed because they persist
(and even become stronger) when the last three nights are removed from the data
set. The ratio 
$f_1/f_7 = 3.500 \pm 0.026$
and is thus exactly equal to a half-integer value. 
This suggests that these frequencies
are related to overtones of each other. On the other hand, the close proximity
of $P_7$ to half a day indicates that this variation may be a data sampling
effect. But note that, considering the error limits, the difference between 
$P_7$ and 0.5~days is rather large
($2.3 \sigma$)\footnote{$2.7 \sigma$, 
if instead of the civil day the sidereal day
is regarded. Systematic measurement errors such as, e.g., an imperfect
extinction correction should be correlated with sidereal time, not civil
time.}. 

In a second approach we prewhitened the combined light curves by first fitting
a sine curve with the period fixed to $P_1$ and then subtracting the fit.
The power spectrum of the resulting data is shown in the third frame of
Fig.~\ref{powerspectra}. As expected, frequencies $f_2$ -- $f_6$ remain
while $f_1$ and its aliases vanish. The complete absence of any residual
signal at $f_1$ means that the waveform of the $P_1$ modulation is very close
to a simple sinusoid. 

Finally, the bottom frame of Fig.~\ref{powerspectra} contains the power
spectrum after removing both, the $P_1$ and $P_2$ modulations. The $f_7$
peak and its aliases remain, while all other previously identified
signals disappear. The power spectrum of a light curve from which also
the $P_7$ signal has been removed does not contain any more signals which
might confidently indicate coherent brightness variations.

\begin{figure}
\parbox[]{0.1cm}{\epsfxsize=14cm\epsfbox{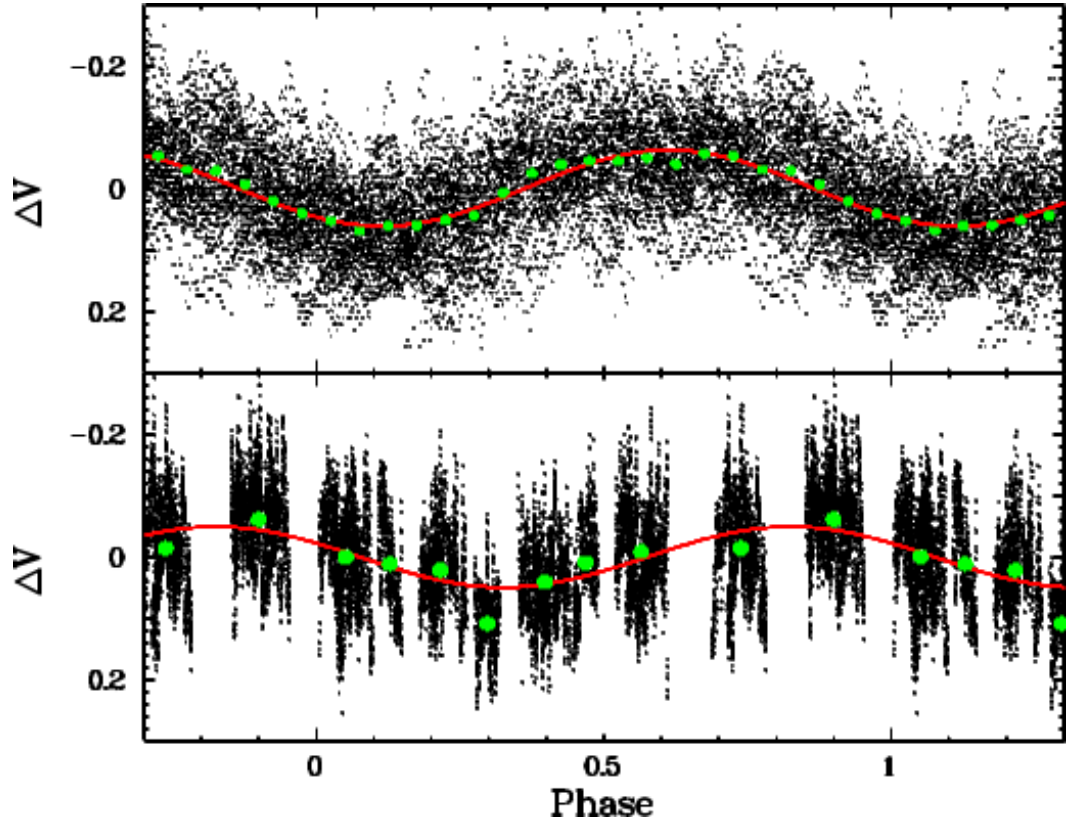}}
      \caption[]{Combined light curves of V603~Aql folded on
                 $P_1$ (left) and $P_2$ (right) after removal of the 
                 $P_2$ and $P_1$ modulation, respectively. The red curves
                 are least squares sine fits to the data. The yellow dots
                 represent binned versions of the folded data in order
                 to highlight eventual deviations of the waveform from
                 a pure sine.}
\label{phase-folded}
\end{figure}

Light curves folded on $P_1$ (after removing the $P_2$ modulation) and
$P_2$ (after removing the $P_1$ modulation) are shown in the left and
right frames, respectively, of Fig.~\ref{phase-folded}. The zero point
of phase was arbitrarily chosen to be HJD 2452800. 
The amplitude of the best fit sine curve to the superhump modulation
($P_1$) is 
$0\hochpunkt{m}0620\pm 0\hochpunkt{m}0007$. 
Thus, the full range of the modulation is twice this value. This is
somewhat more than the range of $\approx$0\hochpunkt{m}08 measured by 
Patterson et al.\ (1993).
The sine fit to the $P_2$ modulation has an amplitude of 
$0\hochpunkt{m}0494 \pm 0\hochpunkt{m}0007$.
  
In order to investigate if the superhump modulation depends on the phase
of the $P_2$ variations the light curves folded on $P_1$ based on only
the nights when on average V603~Aql was brighter and fainter, respectively,
than the mean magnitude over the entire observing run were compared to each
other. No differences in the waveform were detected. The same result was
achieved when selecting light curves on the rising and falling branches,
respectively, of the sine curve in the lower frame of Fig.~\ref{lightcurves}.
Thus, shape and amplitude of the 3\hochpunkt{h}5 signal appear to be
independent from $P_2$.

\section{Discussion}
\label{Discussion}

Observations of superhumps in V603~Aql are not unprecedented (see 
Sect.~\ref{Introduction}). But what about the
$P_2$ modulation? A simple comparison between the orbital frequency 
$f_{\rm orb}$, the superhump frequency $f_{\rm SH} = f_1$ and the frequency $f_2$
shows that the latter is not independent from the others. With the orbital 
period as measured
by Peters \& Thorstensen (2006) 
$f_{\rm orb} - f_{\rm SH} = 0.354 \pm 0.012$.
This is equal to $2 f_2$ within 0.5 times the formal error. The difference
between the orbital and the superhump periods in CVs is well understood as 
being due to the precession of an eccentric accretion disk and the precession
period $P_{\rm prec}$ is given as (see Sect.~\ref{Introduction}) 
$1/P_{\rm prec} = 1/P_{\rm orb} - 1/P_{\rm SH} = f_{\rm orb} - f_{\rm SH}$. 
Thus, $P_2$ can be identified as twice the precession period.

Variations in CVs with positive or negative superhumps occuring on the
precession period are not numerous but they have been observed before. 
Yang et al.\ (2017) compiled 
a list of systems which were reported to exhibit such modulations. We will
concentrate here on positive superhumps, because only these are the topic
of this study and the mechanisms for light modulations on the nodal
precession period of warped accretions disk may be quite different from
those due to apsidal precession of eccentric disks.

In several of the systems enumerated by Yang et al.\ (2017) evidence for
variations with $P_{\rm prec}$ is only marginal or circumstantial, leaving as
the most convincing cases AH~Men, MV~Lyr, TT~Ari and the very object 
investigated here, namely V603~Aql.

In 1993-94 Patterson (1995) observed variations on the beat period between
orbital and superhump periods in AH~Men. Their waveform, however, is
rather complicated. It exhibits three distinct maxima and is thus quite
different from the waveform seen in the right panel of
Fig.~\ref{phase-folded}. 

First indications of variations on $P_{\rm prec}$ in MV~Lyr were reported
by Borisov (1992). While these may be only marginal, confirming observations
by Skillman et al.\ (1995) lend more credibility to their reality. A high
cadence light curve of MV~Lyr spanning hundreds of days with only small 
interruptions has been observed by the Kepler satellite (Scaringi et al. 2012,
Dobrotka et al. 2015). It might be worthwhile to search these data for
similar variations. This, however, is beyond the scope of the present study.

Using spectroscopy, Stanishev et al.\ (2001) found that the asymmetry of the 
H$\alpha$ emission line of TT~Ari is modulated with the precession period. 
This may be explained by the distorted velocity field in a precessing, 
eccentric accretion disk which produces emission lines whose asymmetry is 
modulated with the precession period; an idea first introduced by
Hessman et al. (1992).

Finally, Suleimanov et al.\ (2004) found variations in V603~Aql on the
disk precession period in 2001 and 2002. While this is based on a restricted
data set of only four nightly light curves in 2001 and five in 2002, which
by themselves would not justify a claim for cyclic modulations (see their
Fig.~7), the fact that they can be represented by a sinusoid with a period
independently determined as the beat period of the orbital and an 
observed superhump period gives some confidence in the reality their 
finding.

Compared to all previous claims of variations in CVs on the precession
period of an eccentric accretion disk, the evidence presented here for a 
brightness modulation of V603~Aql not on $P_{\rm prec}$, but on 
$2 \times P_{\rm prec}$, is much stronger. The observations of long
light curves taken in 22 night, 19 of which are consecutive, permit to
determine with high accuracy and confidence the period of superhumps
as well as that of modulations on time scales of days. The latter being
quite precisely equal to twice the beat period between the superhump period 
and the well known orbital period, together with the well established 
understanding for superhumps to arise in a precessing eccentric disk, 
cannot leave any doubt about the connexion between $P_2$ and $P_{\rm prec}$.

However, the immediately following question concerns the origin of the $P_2$
variations. Why does V603~Aql vary on twice the precession period?  
Superhumps occur in eclipsing CVs just as in CVs with a small orbital
inclination. Therefore, the light variations cannot be due to geometrical
effects. It is rather thought that they are caused by the extra tidal 
stresses in the outer, asymmetric disk when its elongated part extends 
towards the secondary. But it is then not clear why the system
brightness should be modulated with the precession period, not to speak
with twice that period, in particular in a system such as V603~Aql where 
the disk is seen amost face-on (Arenas et al. 2000).

Few ideas have been brought forward to explain variations in CV with
$P_{\rm prec}$, and to our knowledge none of them is related to brightness
modulations. Apart from the already mentioned possibility that the
changing view of the velocity field in an eccentric disk may cause
variations in the observed shape of spectral lines (Hessman et al. 1992),
only Patterson et al.\ (2002) provide a clue, interpreting variations in 
the eclipse depth of WZ~Sge on the time scale of days as an effect of disk 
precession in this high inclination system. 

Having thus no explanation for the observed $P_2$ variability we are forced
to leave the solution of this mystery to future studies and as a challenge
for a theoretical interpretation.

\section{Summary}
\label{Summary}

We presented 22 long (mostly $>$5 hours) and high time resolution (35 sec) 
light curves of the 
old nova V603~Aql, 19 of which were taken in consecutive nights. This is 
the longest continuous data train (only interruped during daytime) of this
star ever observed. It is thus not only well suited to study intra-night 
variations occuring on the time scale of hours (such as the well known 
3\hochpunkt{h}5 hour superhump variations), but in particular to investigate
variations on time scales of several days.

These favourable data properties permitted to detect an unexpected feature
in V603~Aql, hitherto unobserved in any cataclysmic variable. The system
exhibits a clear cyclic brightness modulation with a period equal to quite
high a degree to twice the beat period between the orbit and the superhumps.
The latter is routinely interpreted in superhumping CVs as the period of the
apsidal precession of an eccentric accretion disk. 

While indications of varability on the disk precession period have been
observed in a couple of CVs in the past, they have never been nearly as
clear as the variation found in this study on twice that period. It is not
obvious why the brightness of a system should be correlated with the disk
precession period. Geometric effect can probably be ruled out to explain this 
phenomenon, considering that the orbital inclination of V603~Aql 
($13\hoch{o} \pm 2\hoch{o}$) is quite low. And then, why should the brightness
be modulated on {\it twice} the precession period? As yet, we cannot answer
this question.

%

\section*{References}

\begin{description}
\parskip-0.5ex

\item Arenas, J., Catal\'an, M.S., Augusteijn, T., \& Retter, A. 2000,
      MNRAS, 311, 135
\item Belova, A.I., Suleimanov, V.F., Bikmaev, I.F., et al. 2013, Astron.\ 
      Letters, 39, 111
\item Borisov, G.V. 1992, A\&A, 261, 154
\item Bruch, A. 1991, Acta Astron., 41, 101
\item Bruch, A. 1993, 
      A Reference Guide (Astron.\ Inst.\ Univ.\ M\"unster
\item Bruch, A. 2018, New Astron., 58, 53
\item Cropper,M. 1986, MNRAS, 222, 225
\item Dobrotka, A., Mineshige, S., Ness, J.-U. 2015, MNRAS, 447, 316
\item Drechsel, H., Rahe, J., Seward, F.D., \& Wargau, W. 1983, A\&A, 126, 357
\item Drechsel, H., Rahe, J., Wargau, W., \& Krautter, J. 1982, 
      Mitt.\ Astron.\ Ges., 57, 301
\item Eastman, J., Siverd, R., \& Gaudi, B.S. 2010, PASP, 122, 935
\item Haefner, R. 1981, IBVS, 2045
\item Haefner, R., \& Metz, K. 1985, A\&A, 145, 311
\item Hessman, F.V., Mantel, K.-H., Barwig, H., \& Schoembs, R. 1992, 
      A\&A, 263, 147
\item Hollander, A., Kraakman, H., \& van Paradijs, J. 1993, A\&AS, 101, 87
\item Johnson, C.B., Schaefer, B.E., Kroll, P., \& Henden, A.A. 2014, 
      ApJ, 780, L25
\item Kozhevnikov, V.P. 2004, A\&A, 419, 1035
\item Kozhevnikov, V.P. 2007, MNRAS, 378, 955
\item Kozhevnikov, V.P. 2012, New Astron., 17, 38
\item Kraft, R.P. 1964, ApJ 139, 457
\item Lomb, N.R. 1976, Ap\&SS, 39, 447
\item Nogami, D., Masuda, S., Kato, T., \& Hirata, R. 1999, PASJ, 51,115
\item Papadaki, C., Boffin, H.M.J., Stanishev, V., et al.\ 2009, 
      J.\ Astron.\ Data, 15, 1
\item Papadaki, C., Boffin, H.M.J., Sterken C., et al.\ 2006, A\&A, 456, 599
\item Patterson, J. 1981, ApJ Suppl., 45,517
\item Patterson, J. 1995, PASP, 107, 657
\item Patterson, J., Kemp, J., Shambrook, A., et al. 1997, PASP, 109, 1100 
\item Patterson, J., \& Richman, H., 1991, PASP, 103, 735
\item Patterson, J., \& Skillman, D.R. 1994, PASP, 106, 1141
\item Patterson, J., Thomas, G., Skillman, D.R., \& Diaz, M., 1993, 
      ApJS, 83, 235
\item Patterson, J., Thorstensen J.R., Fried, R., et al. 2001, PASP, 113, 72 
\item Peters, C.S., \& Thorstensen, J. 2006, PASP, 118, 687
\item Retter, A., Hellier, C., Augusteijn, T., et al. 2003, MNRAS, 340, 679 
\item Scargle, J.D. 1982, ApJ, 263, 853
\item Scaringi S., K\"ording, E., Uttley, P., et al. 2012, MNRAS, 427, 3396
\item Schwarzenberg-Czerny, A., Udalski, A., \& Monier, R. 1992, ApJ, 401, L19
\item Skillman, D.R., Patterson, J., Thorstensen, J.R. 1995, PASP, 107, 545
\item Smak, J. 2013, Acta Astron., 17, 453
\item Stanishev, V., Kraicheva, Z., \& Genkov, V. 2001, A\&A, 379, 185
\item Strope, R.J, Schaefer, B.E., \& Henden, A.A. 2010, AJ, 140, 134
\item Suleimanov, V., Bikmaev, I., Belyakov, K., et al. 2004,
      Astron.\ Lett., 30, 615
\item Udalski, A., \& Schwarzenberg-Czerny, A. 1989, Acta Astron.,39, 125
\item Yang, M.T.-C., Chou, Y., Ngeowm C.-C., et al. 2017, PASP, 129,4202

\end{description}

\end{document}